\documentstyle[emulateapj,psfig,pstricks,apjfonts]{article}
\newcommand{\etal}{{et~al. }}
\newcommand{\eg}{{\it e.g. }}

\newcommand{\Hi}{\ion{H}{1}}

\newcommand{\snr}{{G21.5--0.9}}

\begin{document}

\title{{\it Chandra} Observations of the Crab-like Supernova Remnant \snr}

\author{Patrick Slane\altaffilmark{1}, 
Yang Chen\altaffilmark{1,2}, 
Norbert S. Schulz\altaffilmark{3},
Frederick D. Seward\altaffilmark{1}, 
John P. Hughes\altaffilmark{4},
and Bryan~M.~Gaensler\altaffilmark{3,5}}

\altaffiltext{1}{Harvard-Smithsonian Center for Astrophysics, 60 Garden Street,
Cambridge, MA 02138}
\altaffiltext{2}{Department of Astronomy, Nanjing University, Nanjing
210093, P. R. China}
\altaffiltext{3}{Center for Space Research, Massachusetts Institute of
Technology, Cambridge, MA 02139}
\altaffiltext{4}{Department of Physics and Astronomy, Rutgers, The State 
University of New Jersey, Piscataway, NJ 08854-8019}
\altaffiltext{5}{Hubble Fellow}

\slugcomment{Submitted for publication to The Astrophysical Journal Letters}

\begin{abstract}
\noindent
{\it Chandra} observations of the Crab-like supernova remnant 
\snr\ reveal a compact
central core and spectral variations indicative of synchrotron burn-off
of higher energy electrons in the inner nebula. The central core is
slightly extended, perhaps indicating the presence of an inner wind-shock
nebula surrounding the pulsar. No pulsations are observed from the
central region, yielding an upper limit of $\sim 40\%$ for the pulsed fraction. 
A faint outer shell may be the first evidence of the 
expanding ejecta and blast wave formed in the initial explosion, indicating
a composite nature for \snr.
\end{abstract}


\section{INTRODUCTION}

Supernova remnants (SNRs) in the ``Crab-like'' or ``plerionic''
class (Weiler \& Panagia 1978) are characterized by compact, filled-center 
radio morphology, with relatively flat spectral index 
($\alpha_r \approx $ 0.0 -- 0.3, where $S_\nu \propto \nu^{-\alpha_r}$).
The common interpretation is that these remnants are pulsar-powered,
although direct evidence of an associated pulsar is often lacking.
Roughly 5\% of the currently cataloged Galactic SNRs (Green 1998) are included
in this class, with another 7--10\% classified as ``composite''
remnants showing plerionic cores accompanied by shells with steeper
radio indices ($\alpha_r \approx$ 0.4 -- 0.7). The shell-like
component is associated with the supernova blast wave and
ejecta.  In X-rays, the plerionic remnants show nonthermal 
spectra characteristic of synchrotron emission while the shell components 
typically display line-dominated thermal emission, although the
shell-type SNRs SN~1006 and G347.3-0.5 are dominated by nonthermal
X-rays (Koyama \etal 1995; Slane \etal 1999).

The absence of any shell or halo of fast-moving ejecta in the Crab
Nebula, 3C58, and other plerionic remnants may be simply the result of a
low ambient density which precludes the formation of a detectable shock.
Evidence of \Hi\ voids around some filled-center
remnants (Wallace, Landecker, \& Taylor 1994) lends support to this
hypothesis, although results for \snr\ in particular are inconclusive. 
However, G74.9+1.2 and G63.7+1.1 are
filled-center SNRs which show no evidence of fast-moving halos of ejecta,
but which appear to be interacting directly with surrounding
material (Wallace, Landecker, Taylor, \& Pineault 1997; Wallace, 
Landecker \& Taylor 1997).
Such an interpretation, if correct, could suggest some neutron stars
form from under-energetic explosions in which very little material is
ejected. Further searches for evidence of extended shell/halo components
in Crab-like remnants is thus of considerable importance.

\snr\ is a compact SNR that exhibits strong linear 
polarization and centrally peaked emission in the radio band (Morsi
\& Reich 1987). The radio spectral index is $\alpha_r \sim 0$ for
$\nu < 32$~GHz, which puts it in the class of Crab-like SNRs. 
A steepening of the spectrum is observed beyond 32~GHz (Salter \etal
1989).
Previous X-ray observations reveal a similar morphology (Becker \& 
Szymkowiak 1981) with a hard X-ray spectrum (Davelaar, Smith, \& Becker 1986).
To date, there has been no detection of a central pulsar
(Frail \& Moffett 1993; Kaspi \etal 1996).

\begin{figure*}[t]
\pspicture(0,11.9)(0,21)
\rput[tl]{0}(0.0,21.7){\epsfxsize=16.4cm
\epsffile{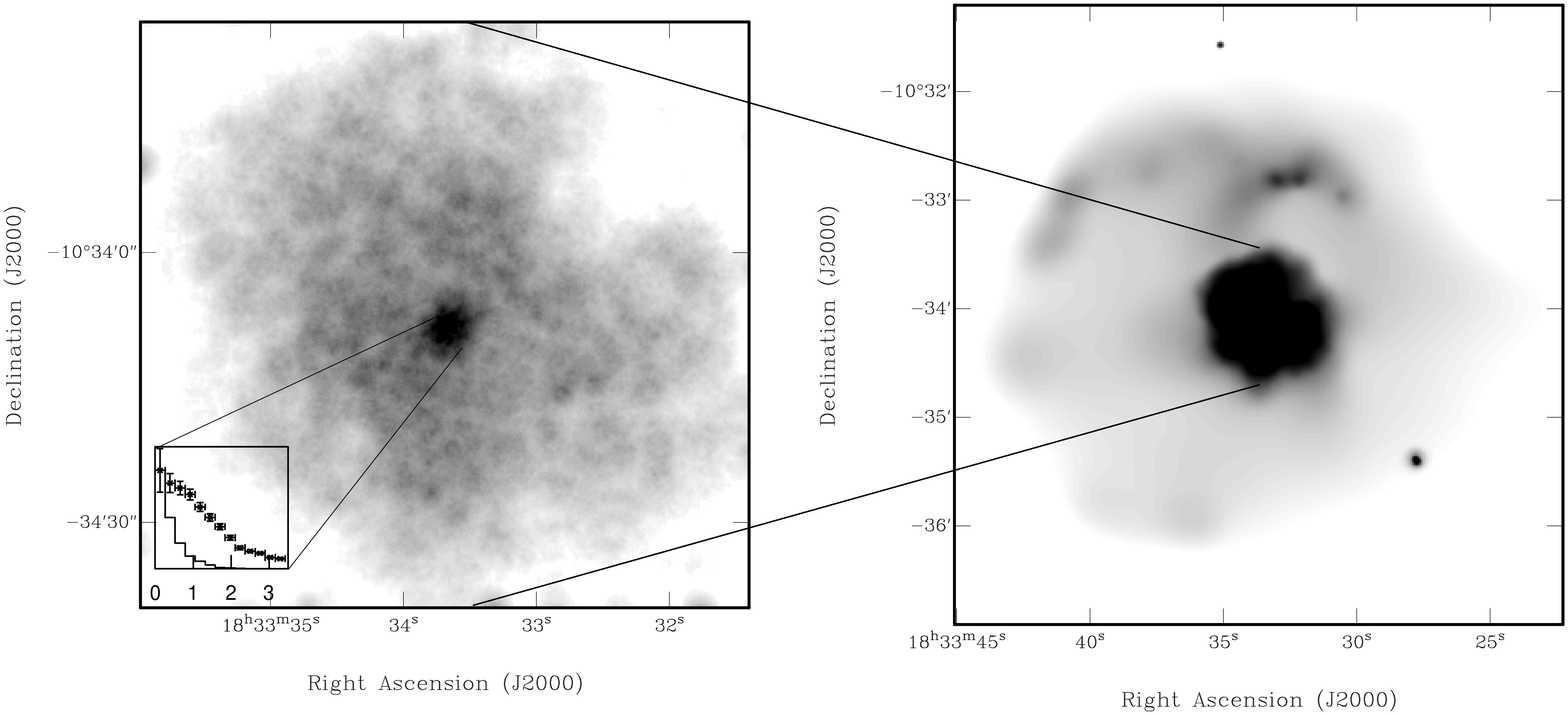}}

\rput[tl]{0}(0,13.5){
\begin{minipage}{16.4cm}
\small\parindent=3.5mm
{\sc Fig.}~1.---
Left: HRC image of the plerionic component of G21.5--0.9. The image 
has been adaptively smoothed with a Gaussian whose width varies inversely
with the number of counts in the image. The inset shows the brightness
profile of the center (points) compared with that for a point source
(histogram). Intensity units are arbitrary, and units on horizontal
scale are arcsec.
Right: ACIS image of G21.5--0.9 
revealing faint halo of emission surrounding the synchrotron core. The image 
has been adaptively smoothed in a manner similar to that used for the
HRC image. The bright source is positionally coincident with the
emission-line star SS 397.
\end{minipage}
}
\endpspicture
\end{figure*}

Neutral hydrogen absorption measurements place \snr\ at a distance of
$\sim 4.8$ kpc (Becker \& Szymkowiak 1981). Throughout
this Letter we assume $d = 5$~kpc and show the scaling of relevant
quantities with $d = 5 d_{5}$~kpc explicitly.

The radio luminosity of \snr\ is $1.8 \times 10^{34} d_{5}^2 {\rm\ erg\
s}^{-1}$ ($\nu < 32$ GHz; Morsi \& Reich 1987),
and it
has a lower $L_x/L_r$ ratio than the Crab; it is a
factor of $\sim 9$ less luminous in the radio and a factor of $\sim 100$ 
less in X-rays. Previous measurements fail to show evidence of any
extended component outside the plerion. 


\section{OBSERVATIONS AND ANALYSIS}

\snr\ has been observed regularly as a calibration target for the {\it Chandra
X-ray Observatory} (Weisskopf, O'Dell, \& van Speybroeck 1996).
Here we concentrate on an early set of observations carried out on
23 Aug 1999 and 25 October 1999.
We select ACIS observations with the remnant placed on the
S3 chip, where the best focus is achieved, and use these for
both spectral and spatial information.
We use observations with the HRC to obtain the highest angular ($\sim 0.5$
arcsec) and time resolution information. The observations used for this study
are summarized in Table 1. 

\begin{center}
{\small
{\bf Table 1: Summary of Observations}\\
\begin{tabular}{llll}\hline\hline
Obs ID & Instrument & Dur. & Date \\ \hline
159 & ACIS (S3) & 17.7 ks & 23 Aug 1999 \\
1230 & ACIS (S3) & 15.8 ks & 23 Aug 1999 \\
1406 & HRC-I & 30 ks & 25 Oct 1999\\ \hline
\end{tabular}
}
\end{center}

In Figure 1 (left) we present the HRC image of the plerion. 
At the center of the X-ray image
is a bright compact region which contains $\sim 8\%$ of the total
flux. This central emission is extended beyond the $\sim 0.5$~arcsec
resolution of the telescope, however.
A comparison of the emission profile with that for a point source at
the same on-axis position
in the HRC reveals an extent of $\sim 2$~arcsec (Fig. 1), 
corresponding to a source size of $\sim 0.05 d_{5}$~pc. 
An investigation of timing data from the
central region yields no significant evidence for pulsations; we derive
an upper limit of $\sim 40\%$ for the pulsed fraction in the HRC 
bandpass ($\sim 0.1$--10~keV) assuming a sinusoidal light curve.

%
%

The next largest structure is the main body of the plerion, which has
a radius of roughly 30 arcsec, with some variation in size -- particularly
in the northwestern region which shows a distinct indentation.
The ACIS data for the plerion reveals a nonthermal spectrum with
energy index $\alpha_x \sim 0.9$ and a column density
$N_H \sim 2 \times 10^{22}{\rm\ cm}^{-2}$. The latter is
consistent with the \Hi\ absorption measurements (Davelaar, Smith,
\& Becker 1986).
Here we have used a radius of 40 arcsec to extract events from the
entire plerion; background was taken from regions of the same detector
well-separated from the remnant.
For comparison, we have used {\it ASCA} data to fit the spectrum and find
excellent agreement with the values derived from the {\it Chandra} data.

The X-ray luminosity
of the plerion is $L_x = 2.1 \times 10^{35} d_{5}^2 {\rm\ erg\ s}^{-1}$.
When the column density is fixed at the value above, we find that
the spectral index varies with radius, as indicated in Figure 2.
This spectral softening with increasing radius, which is also observed
for 3C58 (Torii \etal 2000), is consistent with 
synchrotron burn-off of electrons accelerated in the central regions;
due to the short synchrotron lifetime for higher energy
electrons, only the lower energy component is able to survive to the
outer regions of the plerion. The spectral index of the compact
core is $\alpha_x = 0.5$, similar to that of the Crab pulsar.

Perhaps the most interesting result from the ACIS observations is the
presence of a very faint outer shell or halo in \snr. As
illustrated in Figure 1 (right), the shell radius is roughly
1.2 -- 2 arcmin, corresponding $1.7 - 2.9 d_{5}$~pc. 
The faint shell is incomplete in the southwest, and shows several knots
in the northern region.
The spectrum of this outer region is difficult to quantify at this
early stage of the {\it Chandra} calibration efforts. Using data with a 
combined integration time of 33.5 ks, we obtained four spectra with
a total of $\sim 8700$ counts; these spectra were fit simultaneously.
Combining data from observations with different CCD chips is, as this
stage, not possible. The spectrum, which is relatively hard,
can be adequately described
by either a power law or an optically thin thermal spectrum (e.g.
Raymond \& Smith 1977). The results are listed in Table 2.

\centerline{\psfig{file=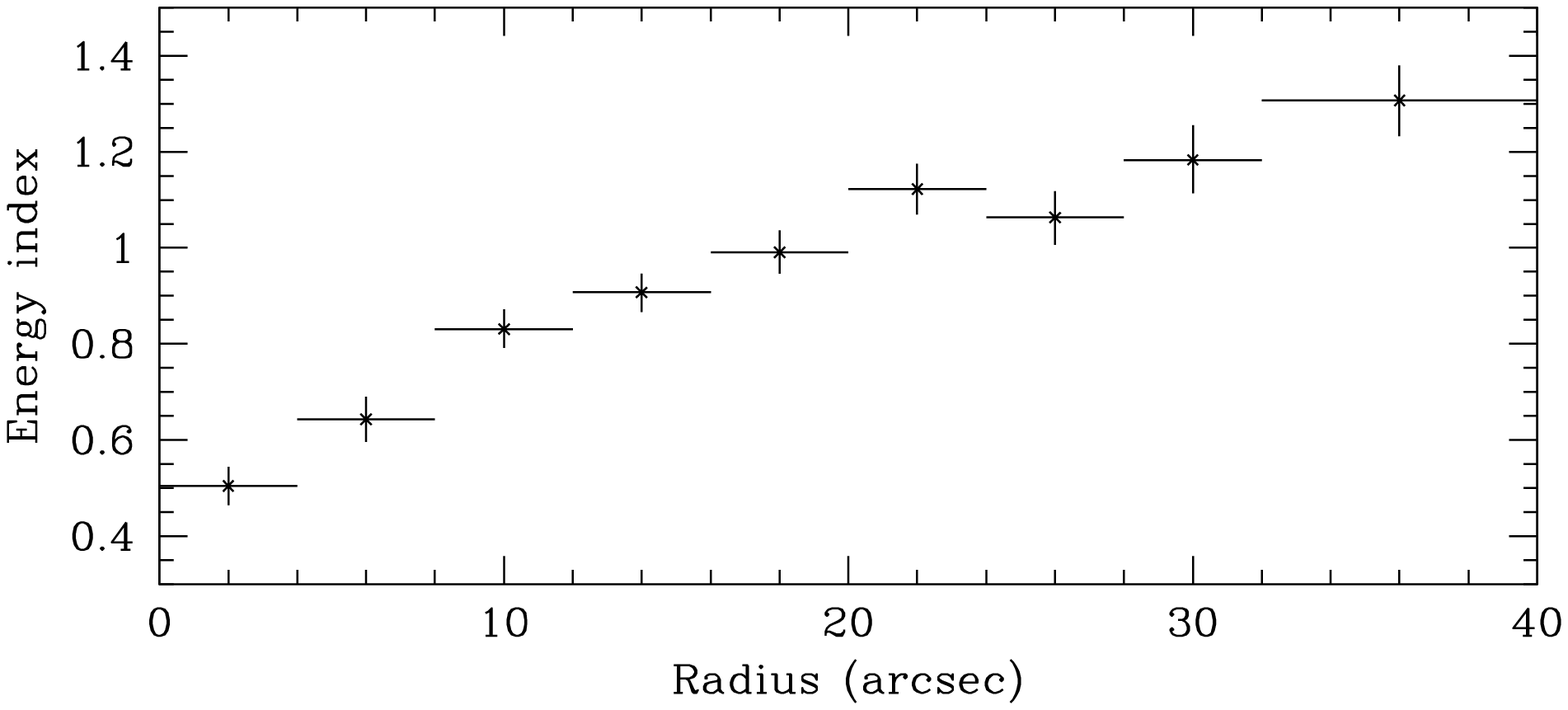,width=7.8cm}}
{\sc Fig.}~2.---
{\small Variation of power law index with radius for the
synchrotron core of G21.5--0.9.}

\begin{figure*}[t]
\begin{center}
\vspace{-0.2in}
{\small
{\bf Table 2: G21.5$-$0.9 Spectral Fits}\\
\begin{tabular}{llllll}\hline\hline
Region & $N_H (\times 10^{22}{\rm\ cm}^{-2})$ & $\alpha_x$ (energy) &
$kT$ (keV) & $F_x ({\rm erg\ cm}^{-2}{\rm\ s}^{-1}$)$^{\rm a}$ & $\chi^2$ (dof) \\ \hline
Plerion$^b$ & $2.24 \pm 0.04$ & $0.91 \pm 0.04$ & -- & $7.0 \times 10^{-11} $
& 2171 (1671)\\
Shell &&&&&\\
(Thermal) & $2.24$ (fixed)& -- & $2.8 \pm 0.2$ & $8.4 \times 10^{-12}$
& 520 (369) \\
(Nonthermal) & $2.24$ (fixed) & $1.56 \pm 0.05$ & -- & $1.2 \times 10^{-11}$
& 379 (369) \\ \hline
\end{tabular}\\
a) unabsorbed flux (0.5 -- 10 keV) \hspace{0.3in}
b) entire synchrotron nebula
}
\vspace{-0.2in}
\end{center}
\end{figure*}

\section{DISCUSSION}

The nonthermal nature of the X-ray emission from the core
of \snr\ indicates the presence of a central pulsar that powers
the synchrotron nebula. X-ray studies of pulsars indicate a
correlation between the total nonthermal X-ray luminosity of the
systems and the spin-down power $\dot E$  of the pulsar (\eg
Seward \& Wang 1988, Becker \& Tr\"{u}mper 1997). Our best fit
for this relation gives $L_x = 3 \times 10^{-11} \dot E^{1.22}$
where we have used 0.1--2.4 keV luminosity values. Using
the luminosity in Table 2, we
infer the presence of a pulsar with $\dot E = 3.5 \times 10^{37}
d_{5}^{1.6} {\rm\ erg\ s}^{-1}$. This is large for the general
pulsar population, but reasonable for a young pulsar associated
with a synchrotron nebula.

Applying standard synchrotron emission calculations (Ginzburg \& 
Syrovatskii 1965) to the radio spectrum from \snr, under the 
assumption of a power law electron distribution (cf. Harrus, Hughes,
\& Slane 1998), 
we derive a nebular magnetic field
$B = 4.4 \times 10^{-4} d_{5}^{-2/7} \theta_{30}^{-6/7}
{\rm\ G}$ and a magnetic pressure $P_B = 7.7 \times 10^{-9} 
d_{5}^{-4/7} \theta_{30}^{-12/7} {\rm\ dyne\ cm}^{-2}$. 

Balancing the ram pressure of a pulsar-driven wind ($\dot E/4 \pi \eta c 
r_w^2$, where the wind covers a fraction $\eta$ of a sphere)
with the nebular magnetic pressure, we estimate a termination
shock radius $r_w \approx 0.04 \eta^{-1/2} d_{5}^{1.1} 
\theta_{30}^{6/7} {\rm\ pc}$. This corresponds to an angular size of 
$\sim 1.5 \eta^{-1/2}$~arcsec for the distance and nebular size estimates 
used above. Interestingly, this is consistent with the apparent extent of 
the central source in \snr\ (Figure 1). It may be that the pulsar itself 
is hidden from our view beneath the shroud of the termination shock.
For the Crab pulsar, radio observations reveal filamentary ``wisps''
(Bietenholz \& Kronberg 1992) thought to be produced at this boundary
zone where the pulsar wind begins decelerating through interactions
with the surrounding medium (Rees \& Gunn 1974). Similar evidence of
a radio wisp is observed in 3C 58 (Frail \& Moffett 1993), and 
the separation of the feature from the compact X-ray source observed
in this SNR (Helfand, Becker, \& White 1995) appears compatible with 
calculations of the expected termination shock distance (Torii \etal 2000).

The extrapolation of the X-ray spectrum of the plerion as a whole to 
infrared frequencies is inconsistent with the measured ISO flux 
(Gallant \& Tuffs 1998); the measured IR flux exceeds the prediction.
This curious result may result from oversimplification
of the X-ray spectrum, however. As shown in Figure 2, the spectral
index increases with radius. A value of $\alpha_x \sim 1.0$, 
compatible with the outer regions of the nebula, provides good agreement upon 
extrapolation to the IR band. 
Alternatively, this may be indicative of the time evolution of the
energy input into the nebula from the central source.
We note that the X-ray core has no
observed counterpart in the radio or infrared bands.

The faint, extended halo of emission from \snr\ presumably represents
the contribution from the SNR ejecta, with the brighter clumps in
the northwest being associated with regions where the ejecta have
encountered material of higher density. Faint arcs discernible in
Figure 1 could be indicative of a reverse shock being driven into
the expanding ejecta. 
Alternatively, the X-ray emission from the diffuse halo could be 
dominated by synchrotron emission from relativistic electrons accelerated 
at the shock front as proposed for SN~1006 and G347.3--0.5.

The presence of an extended X-ray halo surrounding the Crab-like core
of G21.5--0.9 raises the question of whether there is also an extended
radio halo corresponding to the shell morphology seen in the vast
majority of radio SNRs. We have examined archival Very
Large Array (VLA) data, and find no radio shell around G21.5--0.9 down
to a 1~GHz surface brightness limit (1$\sigma$) of $\Sigma =
4\times10^{-21}$~W~m$^{-2}$~Hz$^{-1}$~sr$^{-1}$.  While this is
sufficient to detect most young SNRs,
at least one Crab-like source, G322.5--0.1, has a shell component which 
would not be seen down to this limit (Whiteoak 1992). 

If the extended halo truly represents the fast-moving ejecta, we
would expect the spectrum to reveal characteristic emission line features
from the hot gas. As summarized in Table 2, the best-fit spectrum is
an absorbed power law. 
While we present results with the column density fixed to that obtained
for the plerion, we note that slightly better spectral fits are obtained
for a somewhat lower value of $1.8 \times 10^{22}{\rm\ cm}^{-2}$.
A model for
thermal emission (Raymond \& Smith 1977) provides an adequate fit
(though inferior to the power law) with a temperature of $2.8$ keV.
Based upon the thermal model, the mean density of the emitting material is 
$n_H = 1.1 \phi_{1.5}^{-3/2} d_5^{-1/2} f^{-1/2}{\rm\ cm}^{-3}$
where $\phi_{1.5}$ is the angular radius of the shell in units of
1.5~arcmin, and $f$ is the fraction of the spherical volume comprising
the X-ray emission. The mass of the emitting material is then
$M_x = 1.2 \phi_{1.5}^{3/2} d_5^{5/2} f^{1/2} M_\odot$. With such a
small amount of swept up material, \snr\ is still dynamically young.

If the extended emission is nonthermal, then \snr\ joins SN~1006 and 
G347.3--0.5 as remnants whose shell emission is dominated by nonthermal 
processes in X-rays. The absence of an observed radio shell is
consistent with the low radio surface brightness of these other remnants.
More importantly, since the plerionic core in \snr\ implies a massive
progenitor, while SN~1006 is the result of a Type Ia explosion, 
the addition of \snr\ to this class would imply that the nonthermal
X-ray shells are related more to environment than to progenitor type.


\section{CONCLUSIONS}
The early {\it Chandra} observations of \snr\ have revealed new information
on all spatial scales. The variation in spectral index with radius in
the plerion supports the view that a central object injects energetic
electrons into the surrounding region, and that the lower energy
particles, whose synchrotron radiation lifetimes are longest, 
predominate at the outskirts of the nebula. The center of the plerion
shows a compact core which presumably contains a neutron star that powers
the system, but this core is not point-like; an extended
shroud, possibly associated with the pulsar wind termination shock,
apparently hides the pulsar itself. No pulsations are detected from
the central region.

On the largest scales, a faint halo of emission is observed, with
some evidence of arcs and clump-like features. This may correspond
to the contribution from the fast-moving ejecta, although the
spectral information is too sparse to quantify the nature of the
emission in detail. If this component does indeed correspond to
the ejecta component, then \snr\ does not fall into the proposed
category of low explosion energy remnants with no ejecta proposed
as possible progenitors to the class of plerionic SNRs. Deeper
observations of the remnant are needed to better constrain this
extended component. Upcoming studies with XMM have an excellent
chance of clarifying this situation.

\acknowledgments

This work was supported in part by the National Aeronautics and
Space Administration through contract NAS8-39073 (P.S.),
and grant NAG5-6420 (J.P.H). Y.C. acknowledges
support from the Huaying Cultural and Educational Foundation.
B.M.G. acknowledges the support of NASA through Hubble Fellowship grant
HF-01107.01-98A awarded by STScI, which is operated by AURA Inc.
for NASA under contract NAS 5--26555.

\end{document}